\documentclass[final,english]{bullsrsl}[2022/06/15]
\usepackage[latin1]{inputenc}
\usepackage[T1]{fontenc}
\usepackage{textcomp}
\usepackage{natbib} 
\usepackage{graphicx}
\usepackage[normalem]{ulem}
\usepackage{xcolor}

\begin{document}
\title{The PRL 2.5m Telescope and its First Light Instruments: FOC \& PARAS-2}

\author[corresponding]{Abhijit}{ Chakraborty}
\author[]{Kapil Kumar }{Bharadwaj}
\author[]{J.S.S.V.Prasad }{Neelam }
\author[]{Rishikesh}{Sharma}
\author[]{Kevikumar A.}{Lad}
\author[]{Ashirbad}{Nayak}
\author[]{Nikitha}{Jithendran}
\author[]{Vishal}{Joshi}
\author[]{Vivek Kumar}{Mishra}
\author[]{Nafees}{Ahmed}
\affiliation[]{Astronomy and Astrophysics division, Physical Research Laboratory, Navarangpura, Ahmedabad, 380009, India}
\correspondance{abhijit@prl.res.in}
\date{15th Jan 2024}
\maketitle{\centering\textit{\textcolor{blue}{(Accepted for publication in the Bulletin of Liege Royal Society of Sciences, Volume 93, 2024)}}}

\begin{abstract}
We present here the information on the design and performance of the recently commissioned 2.5-meter telescope at the PRL Mount Abu Observatory, located at Gurushikhar, Mount Abu, India. The telescope has been successfully installed at the site, and the Site Acceptance Test (SAT) was completed in October 2022. It is a highly advanced telescope in India, featuring the Ritchey-Chr$\acute{e}$tien optical configuration with primary mirror active optics, tip-tilt on side-port, and wave front correction sensors. Along with the telescope, its two first light instruments  {namely Faint Object Camera (FOC) and PARAS-2} were also integrated and attached with it in the June 2022. {FOC is a} camera that uses a 4096 X 4112 pixels detector SDSS type filters  with enhanced transmission and known as u', g', r', i', z'. It has a limiting magnitude of 21 mag in 10 minutes exposure in the r'-band. The other first light instrument PARAS-2 is a state-of-the-art high-resolution fiber-fed spectrograph operates in 380-690 nm wave-band, aimed to unveil the super-Earth like worlds. The spectrograph works at a resolution of $\sim$107,000, making it the highest-resolution spectrograph in Asia to date, which is under  {ultra}-stable temperature and pressure environment, at 22.5 $\pm$ 0.001 $^{\circ}$C and 0.005 $\pm$ 0.0005 mbar, respectively. Initial calibration tests of the spectrograph using a Uranium Argon Hollow Cathode Lamp (UAr HCL) have yielded intrinsic instrumental RV stability down to 30 cm s$^{-1}$.
\end{abstract}
\section{Introduction}
\noindent  {Physical Research Laboratory (PRL)} has been successfully operating its 1.2m telescope from Mt. Abu since late 1994, generating significant scientific contributions in the field of Astronomy \& Astrophysics (\citet{telescope, Anandarao2010} and references therein). In 2008, PRL initiated its exoplanet science program \citep{Chakraborty2008} for the first time in India with the development of a high-resolution spectrograph PARAS (PRL Advanced Radial-velocity Abu-sky Search, R $\sim$ 67000), coupled with the 1.2m telescope at PRL Mount Abu Observatory \citep{Chakraborty2014}.  {However in early 2014, PRL Astrophysicists recognized the need for a larger aperture telescope specifically tailored to PRL's dedicated exoplanetary science  {and target of opportunities (TOO)} programs. The primary aim was to enhance the capability to detect smaller exoplanets, particularly those classified as super-Earths, and to facilitate research in transient phenomena such as gamma-ray bursts (GRBs), novae, and supernovae.} Given PRL's strong reputation in both national and international exoplanetary sciences \citep{Chakraborty2014} and novae sciences \citep{Banerjee2002, AshokN2003}, the decision was made to initiate the project for a 2.5m telescope under the leadership of Project Director AC. In February 2015, PRL solidified its commitment by signing a contract with {M/s.} Advanced Mechanical and Optical Systems (AMOS) in {Leige}, Belgium, marking a significant milestone in the project advancement. Based on the technical specifications designed by the PRL project team, the kick-off meeting between PRL and AMOS was held in mid-July 2016 \citep{PirnayO2018}. Subsequently, Preliminary Design Review (PDR), Critical Design Review (CDR) and Final Design Review (FDR) took place in November 2016, March 2017 and September 2017, respectively. The Factory Acceptance Test (FAT) was conducted in September 2020 and Site Acceptance Test (SAT) was concluded by October 2022. {Mr. Olivier Pirany was the Project Manager for this project from AMOS side.}

\noindent During the same period, PRL also undertook the design and development of two first light instruments for the newly planned telescope. One of these instruments is the FOC,  {designed primarily for detailed scientific observations and study of Transient Astrophysical Phenomena such as Novae, Supernovae, and GRBs.} The FOC exhibits a high sensitivity, capable of detecting objects down to the 21$^{st}$ magnitude in the r'-band with just a 10-minute exposure time. The camera provides a high spatial resolution with a fine plate scale of 0.15" pixel$^{-1}$, and offers a total FOV of 10 arcmins$^{2}$.

\noindent Another first light instrument is the PARAS-2 spectrograph   \citep{Chakraborty2018}, which is a state-of-the-art high-resolution spectrograph   {with R $\sim$ 107,000.} The motivation behind development of the spectrograph stems from the significantly increasing population of small-sized exoplanetary candidates revealed through dedicated space-based missions like K2 \citep{Howell_2014}, TESS \citep{Ricker2015}, PLATO \citep{Rauer2014} , and CHEOPS \citep{Benz2020} as well as the limited availability of high-precise RV spectrographs (R $\geq$ 100,000) worldwide, such as HARPS \citep{Mayor2003}, ESPRESSO \citep{Pepe2010}, NEID \citep{Schwab2016}, HPF \citep{Mahadevan2012,Mahadevan2014} and CARMENES \citep{Quirrenbach2016}.  {A high-precision RV spectrographs (R$\geq$100,000) on 2.5m to 4m telescopes provides an efficient way to conduct exoplanet research, addressing challenges in obtaining time on larger telescopes} \citep{Chakraborty2018}. PARAS-2 aimed to achieve sub-m s$^{-1}$ RV precision, which is necessary to detect and characterize super-Earths or massive planets around G and K dwarf stars. The project to develop PARAS-2 commenced in the mid-2018, and the spectrograph was successfully installed at the telescope site in the mid-2022. PARAS-2 is meticulously maintained within a highly controlled environment, ensuring a precise temperature of 22.5 $\pm$ 0.001$^{\circ}$C and pressure of 0.005 $\pm$ 0.0005 mbar. This newly envisaged fiber-fed spectrograph has been integrated with the recently installed 2.5m telescope at the PRL Mount Abu observatory. The larger aperture of the 2.5m telescope allows for the observation of fainter objects, expanding the capabilities of exoplanetary research.
In this paper, we will briefly discuss the telescope design, development and its on-sky performance results. We will also provide a overview of the development of FOC and the PARAS-2 spectrograph at PRL.

\section{The 2.5m Telescope}\label{obs}
\noindent {The PRL 2.5m telescope project is a state-of-the-art technological development at PRL, India in collaboration with AMOS, Belgium.} The site for the 2.5m telescope at Gurushikhar is located at 24$^{\circ}$ 35' 30" N and 72$^{\circ}$ 46' 30" E, in the Aravalli Ranges at a height of $\sim$1700 m and, around 20 km from Mount Abu.  {Extensive site studies by \citet{site_2, site_1} reveal that Gurushikhar is one of the best astronomical sites in India which provides ideal conditions for astronomical observations in the optical and near-infrared wavelength bands with $\sim$65\% ($\sim$240) of clear nights during a year.} The launch of the new telescope at PRL Mount Abu Observatory intends to improve PRL astronomers' access to medium-sized optical telescopes significantly.

\subsection{Telescope general specifications}
\noindent General specifications of the PRL 2.5m telescope are listed in Table \ref{t2}. \\
\begin{table}[!h]
\caption{PRL 2.5m Telescope General Specifications}\label{t2}
\begin{center}
    \begin{tabular}{| p{1.5cm} | p{4.5cm}| p{6cm} | }
    \hline
    \textbf{Sl. No} & \textbf{Parameters} & \textbf{Details}  \\
    \hline
    1 & Aperture &  2.56 m diameter \\
    \hline
    2 & Configuration &  Ritchey-Chr$\acute{e}$tien (RC) \\
    \hline
    3 & Mount & Alt-Azimuth \\
    \hline
    4 & Primary mirror focal-ratio & F/2 \\
    \hline
    5 & Primary mirror thickness & 120 mm \\
    \hline
    6 & Operating wavelength & 370 nm to 4000 nm \\
    \hline
    7 & Effective focal-ratio & F/8\\
    \hline
    8 & Effective Focal length &  20 m \\
    \hline
    9 & Plate scale & 10.313" / mm \\
    \hline
    10 & FOV (un-vignetted) & 25' diameter at main port, 10' diameter at two side ports\\
    \hline
\end{tabular}
\end{center}
\end{table}

 \noindent The design of the main tube is a Ritchey-Chr$\acute{e}$tien (RC) configuration, i.e., a hyperbolic concave primary mirror (M1) and a hyperbolic convex secondary mirror (M2) sharing a common conical focus.  The focal plane is 2.1 m behind the vertex of the M1.  The meniscus primary mirror of aperture size of 2.5m has advanced active optics support system with 42 axial and 16 lateral pneumatic actuators. The active optics through the wave front sensor (WFS) maintain the surface profile of the mirror during observations. This mitigates the first-order astigmatism, which enhances the overall quality of the star image PSF. The secondary mirror is mounted over the hexapod through the whiffle-tree support system and has five degrees of freedom (tip, tilt, focus) to correct for optical aberrations. There are two separate M3 mirrors to direct the central light to two side ports respectively. One of the M3 mirrors is equipped with a tip-tilt mechanism which will compensate for the first-order atmospheric seeing conditions and thus will improve the star image sharpness giving more explicit and crisper images  {(see Sec~\ref{paras2_cass})}. The details about the telescope design can be found in \citet{PirnayO2018}. \\\\

\begin{figure}[!ht]
\begin{minipage}{9.0cm}
\includegraphics[width=8.5cm,trim=0 0 0 0]{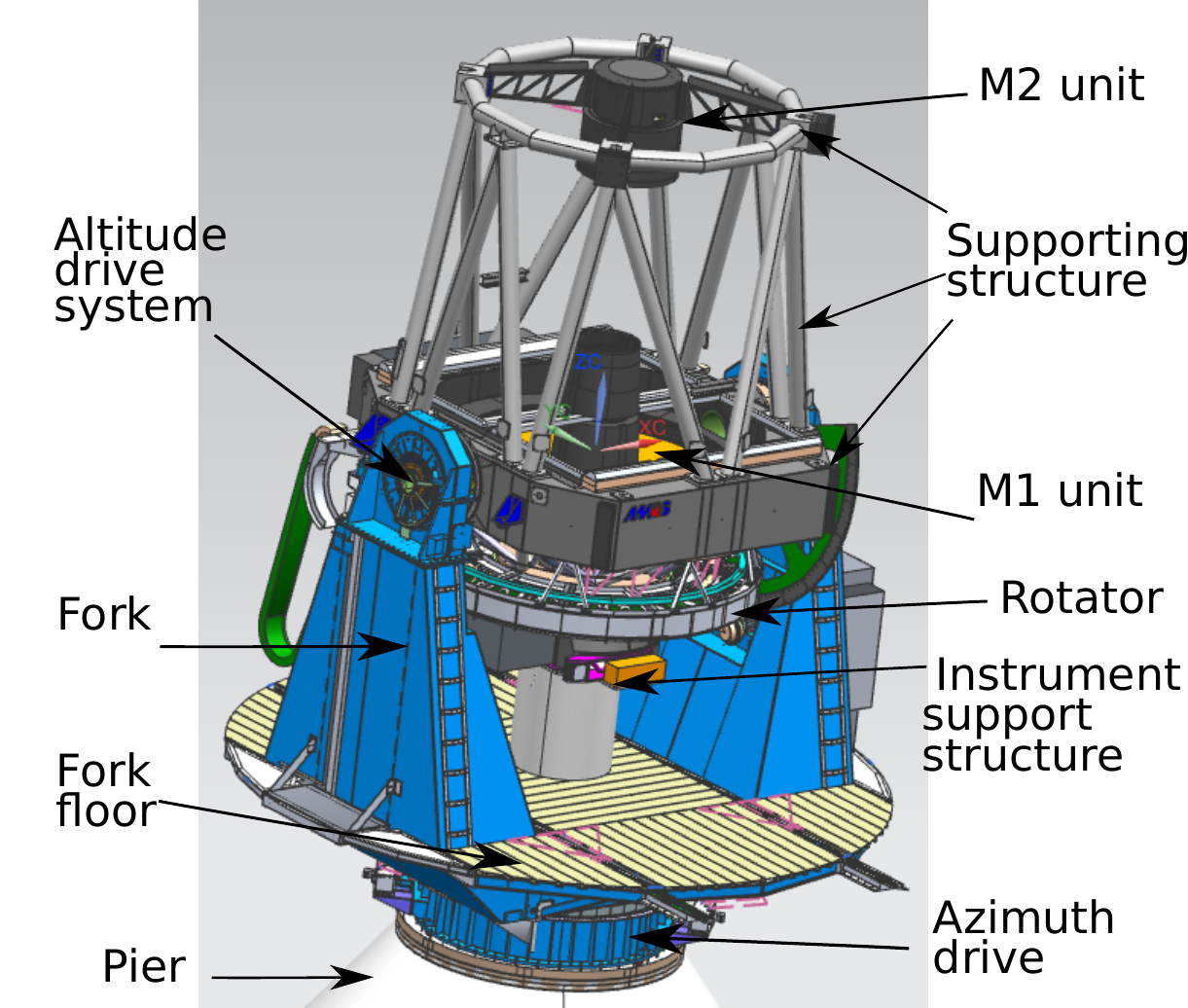}
\end{minipage}
\begin{minipage}{7.0cm}
\includegraphics[width=6cm,trim=0 0 100 0]{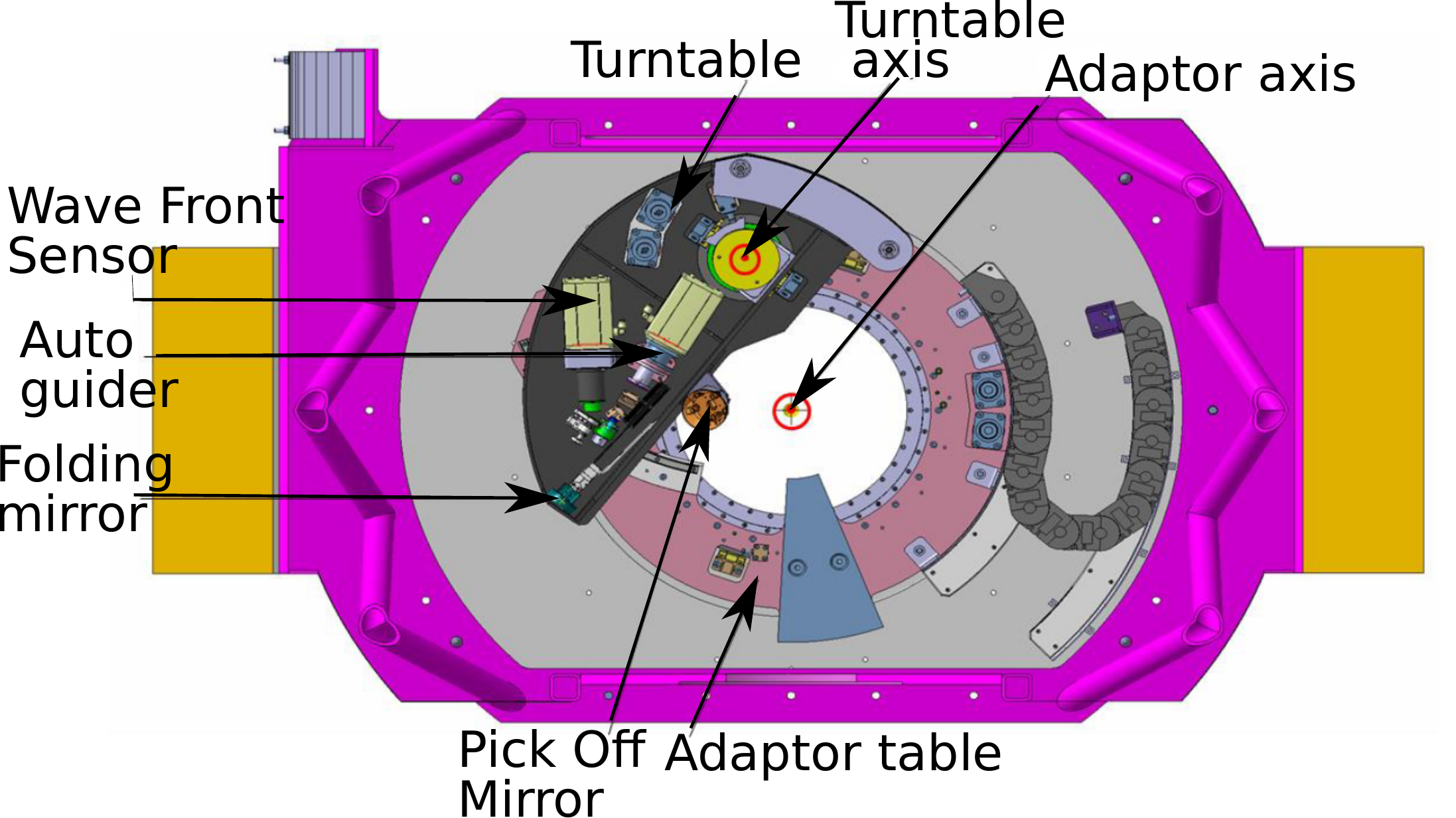}
\includegraphics[width=6cm,trim=0 0 100 0]{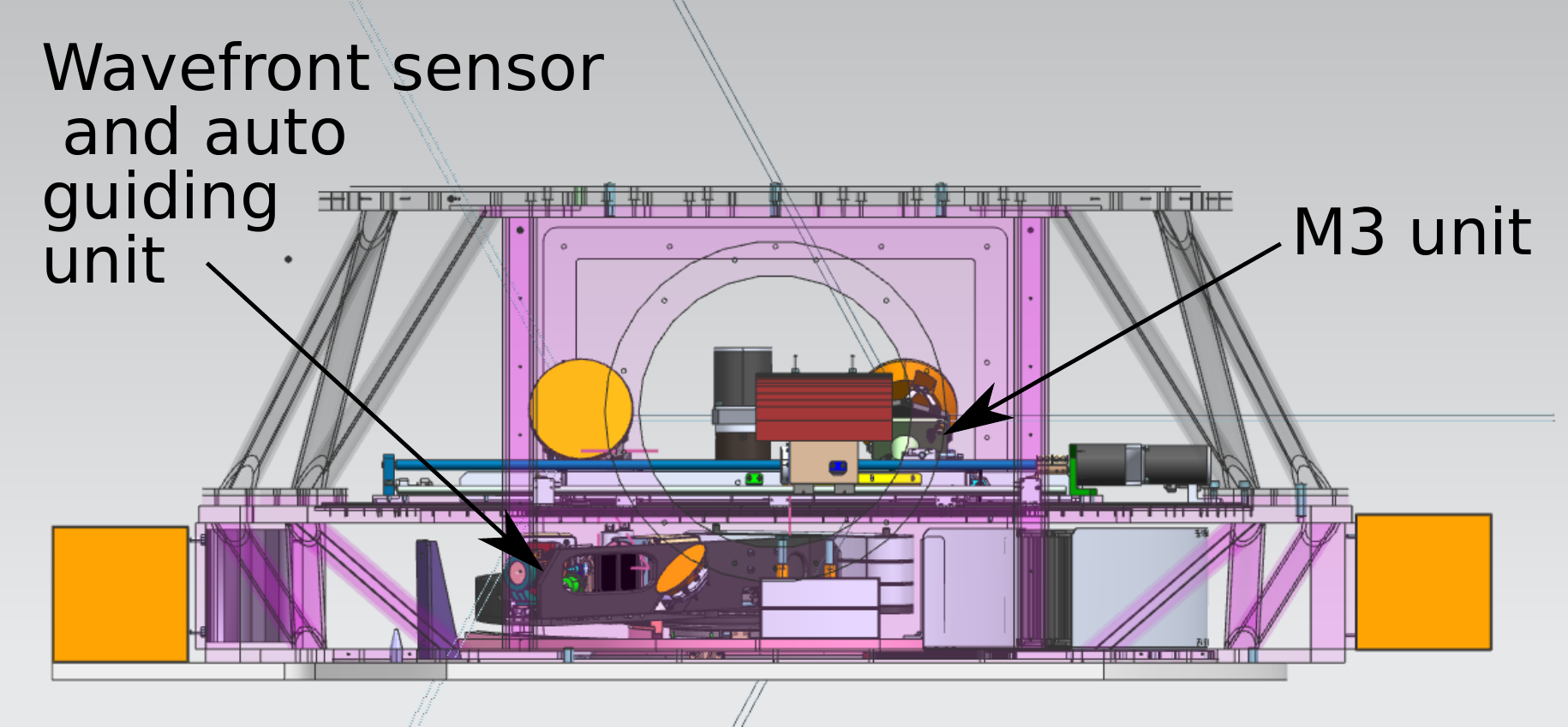}
\end{minipage}
\bigskip
\begin{minipage}{12cm}
\caption{ {The CAD model of the PRL 2.5m telescope and its various subsystems. (Credit - AMOS)}}
\label{fig:1}
\end{minipage}
\end{figure}

\subsection{ {Final Telescope Performance observed during SAT}}
\noindent Initially the Factory Acceptance Test (FAT) was conducted at AMOS, Belgium, in the last week of September 2020, during which various telescope parameters were tested and demonstrated as per the contract conditions between PRL and AMOS before it was shipped to India. The telescope was tested for pointing and tracking accuracy in open and closed loops, and the image quality was evaluated. Absolute Pointing accuracy was measured and found to be better than 1.24 arcsec rms, tracking accuracy was measured as less than 0.22 arcsec rms in an open loop for a short time, and better than 0.16 arcsec rms in a closed loop for long exposure. Further, the star FWHM was found to be 0.33 arcsec using Lucky Imaging. This makes the 2.5m telescope as one of the most precisely polished and surface-figured telescope in the country, with an overall wave front error of around 70 nm RMS. The surface accuracy of the primary mirror was measured to be $\sim$25nm rms at the AMOS facility in Belgium.

\noindent In the first week of February 2021, the telescope arrived at the telescope site at PRL Mount Abu Observatory, Gurushikhar, Mount Abu, India. Installation at the site was planned between February 2021 and the first week of May 2021. {However due to COVID-19 pandemic situation and monsoon in India, the telescope installation started in the late October 2021 at the Gurushikhar site in Mount Abu. The telescope installation was completed by the end of May 2022. After the Indian monsoon, SAT was conducted between October 16 and October 21, 2022.} The main findings from the SAT for the 2.5-meter telescope are listed in Table \ref{t1}. The detailed description of the telescope first light and the performance results can be found in \citet{Bastin2022}. 

\begin{table}[!h]
\caption{Site Acceptance Test (SAT) results of the PRL 2.5m telescope
}\label{t1}
\begin{center}
    \begin{tabular}{| p{0.75cm} | p{3cm}| p{5.5cm} | p{4cm} |}
    \hline
    \textbf{Sl. No} & \textbf{Parameters} & \textbf{Specifications} & \textbf{Performance} \\
    \hline
    \multicolumn{4}{|c|}{\textbf{\centering{Pointing Accuracy}}}\\
    \hline
    1 & Absolute & 2 arcsec RMS & \textbf{1.88 arcsec RMS}\\
    \hline
    2 & Differential & 0.5 arcsec RMS & \textbf{0.45 arcsec RMS}\\
    \hline
    \multicolumn{4}{|c|}{\textbf{Tracking Accuracy}}\\
    \hline
    3 & Open-Loop 1 min & 0.2 arcsec RMS & \textbf{0.13 arcsec RMS}\\
    \hline
    4 & Open-Loop 10 min & 0.5 arcsec RMS & \textbf{0.14 arcsec RMS}\\
    \hline
    5 & Closed-Loop 60 min & 0.2 arcsec RMS & \textbf{0.19 arcsec RMS}\\
    \hline
    \multicolumn{4}{|c|}{\textbf{Image Quality}}\\
    \hline
    6 & Wave Front Error & 70 nm RMS + Seeing degradation & \textbf{64 nm RMS with Shack Hartmann WFS}\\
    \hline
    7 & Full Width at Half Maximum(FWHM) & 0.35 nm RMS + Seeing degradation & \textbf{0.18 arcsec (Lucky Imaging)}\\
    \hline
\end{tabular}
\end{center}
\end{table}

\begin{figure}[!ht]
\begin{minipage}{7.0cm}
\includegraphics[width=7.0cm,height=9.5cm]{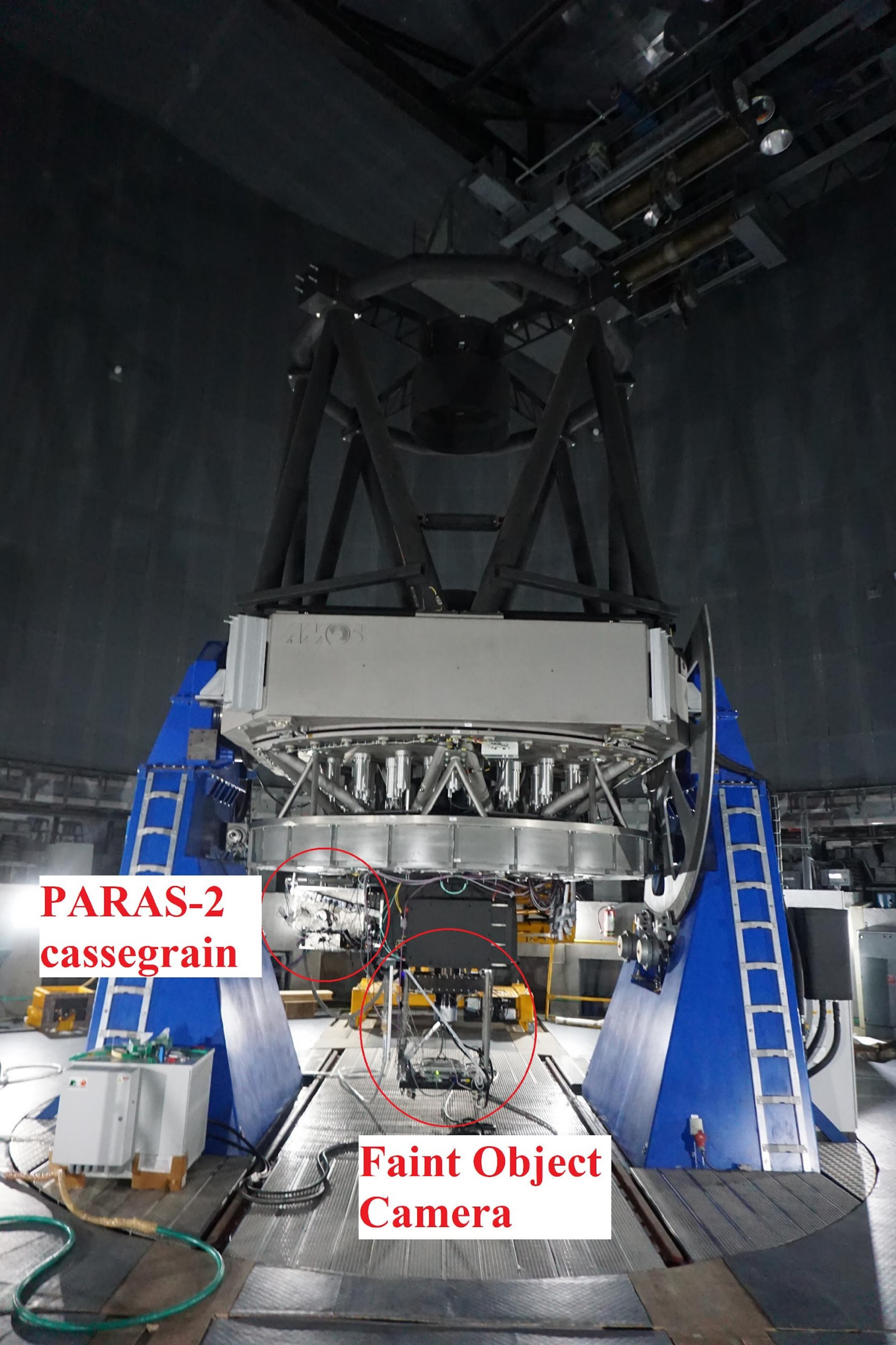}
\end{minipage}
\begin{minipage}{9cm}
\includegraphics[width=7.5cm,height=4.7cm]{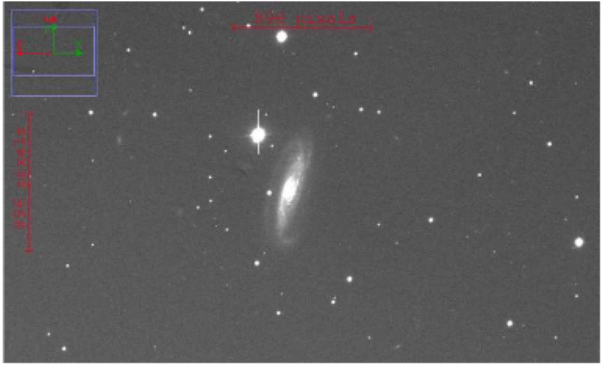}
\includegraphics[width=7.5cm,height=4.7cm]{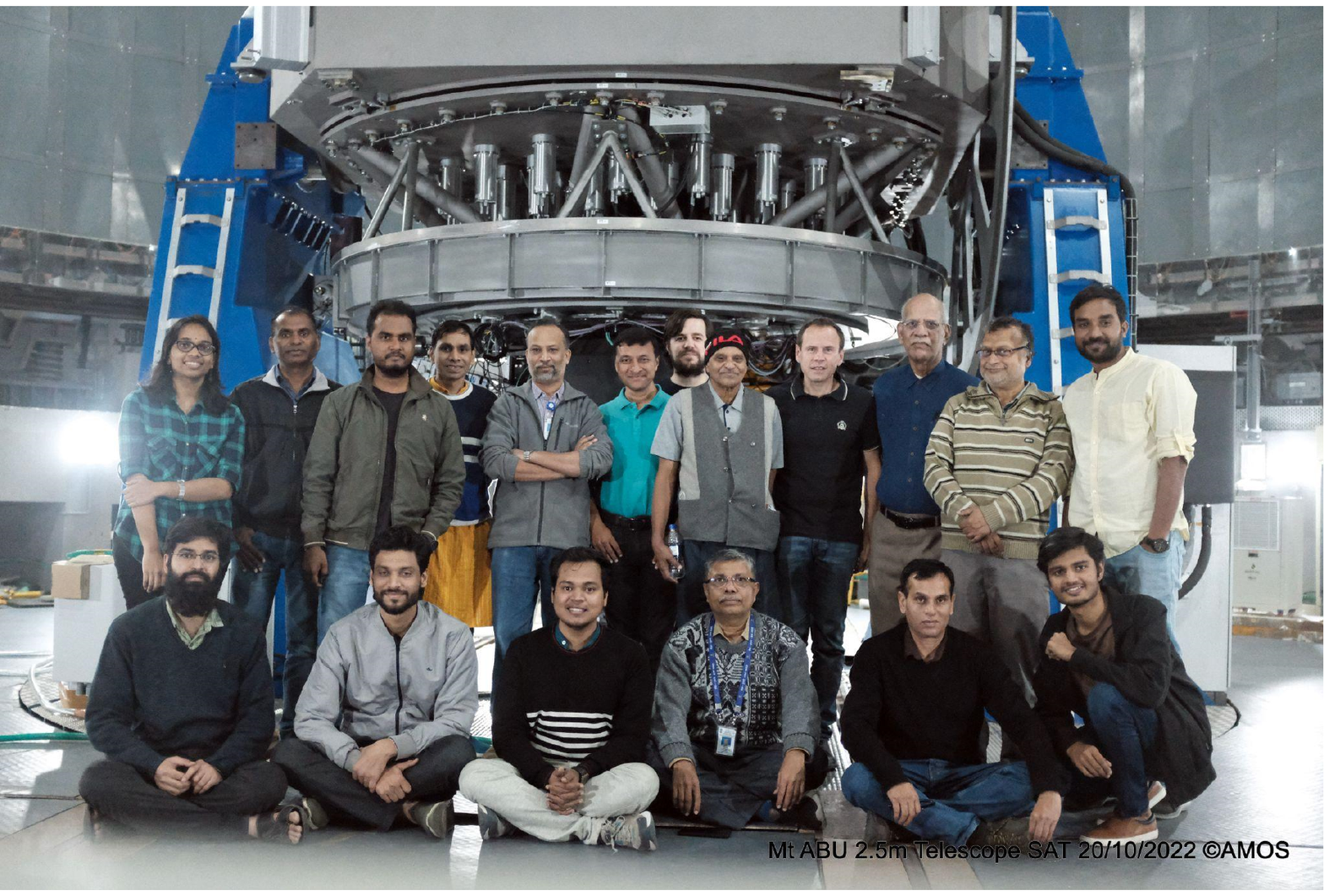}
\end{minipage}
\bigskip
\begin{minipage}{12cm}
\centering
\caption{{Left:- The fully assembled PRL 2.5m telescope at the PRL Mount Abu Observatory, Mount Abu, India. The first light instruments, FOC and the PARAS-2 Cassegrain unit are attached to the main port, and to side port\#1, respectively as circled in the image.}
{Right:- The r'-band image of NGC 295, located $\sim$250 million light-years away, captured using FOC on 16 October 2022. Exposure time =  300 sec ; Seeing =1.1". The faintest star in the field is $\sim$20.5 mag, and the sky background is 21 mag per arcsec. Moon was 50 degrees away and was at 60\% of its brightness (top). PRL 2.5m telescope team after successful conclusion of the SAT (bottom).}}
\label{fig:2}
\end{minipage}
\end{figure}

\begin{figure}[!ht]
\centering
\includegraphics[width=14cm]{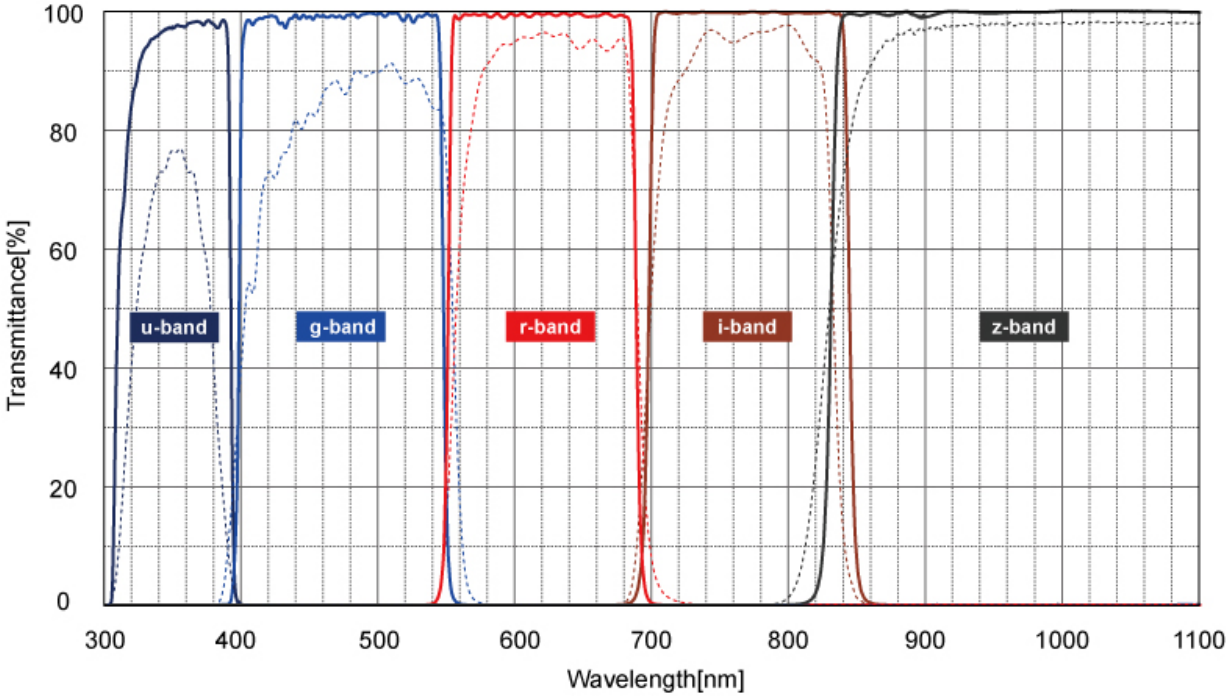}
\bigskip
\begin{minipage}{12cm}
\caption{Comparison of Transmission of traditional SDSS filters and FOC filters. The dashed lines are traditional standard SDSS ugriz bands whereas solid lines are FOC filters. The solid lines show the transmission response curves of u',g',r',i', z' filter sets used in FOC. }
\label{fig:3}
\end{minipage}
\end{figure}

\section{ {First Light by Faint Object Camera (FOC)}}
\noindent The Faint Object Camera (FOC) is a key instrument for the 2.5m Telescope at Mount Abu Observatory, which has been in use since October 2022, allowing for the acquisition of images and photometry of celestial objects in the {various} wave-bands. This instrument also helped us to verify the telescope performance results established during the telescope SAT. FOC offers 5 different filters SDSS type with improved transmission compared to traditional SDSS filters (see figure~\ref{fig:3}) covering wavelength between 350 nm to 1000 nm and is addressed as u', g', r', i', and z'. The same filters are used with GTC OSIRIS located at Observatorio del Roque de Los Muchachos in La Palma, Spain \citep{GTC_filters}.

\begin{figure}[!ht]
\centering
\includegraphics[width=9cm, height=7.3cm]{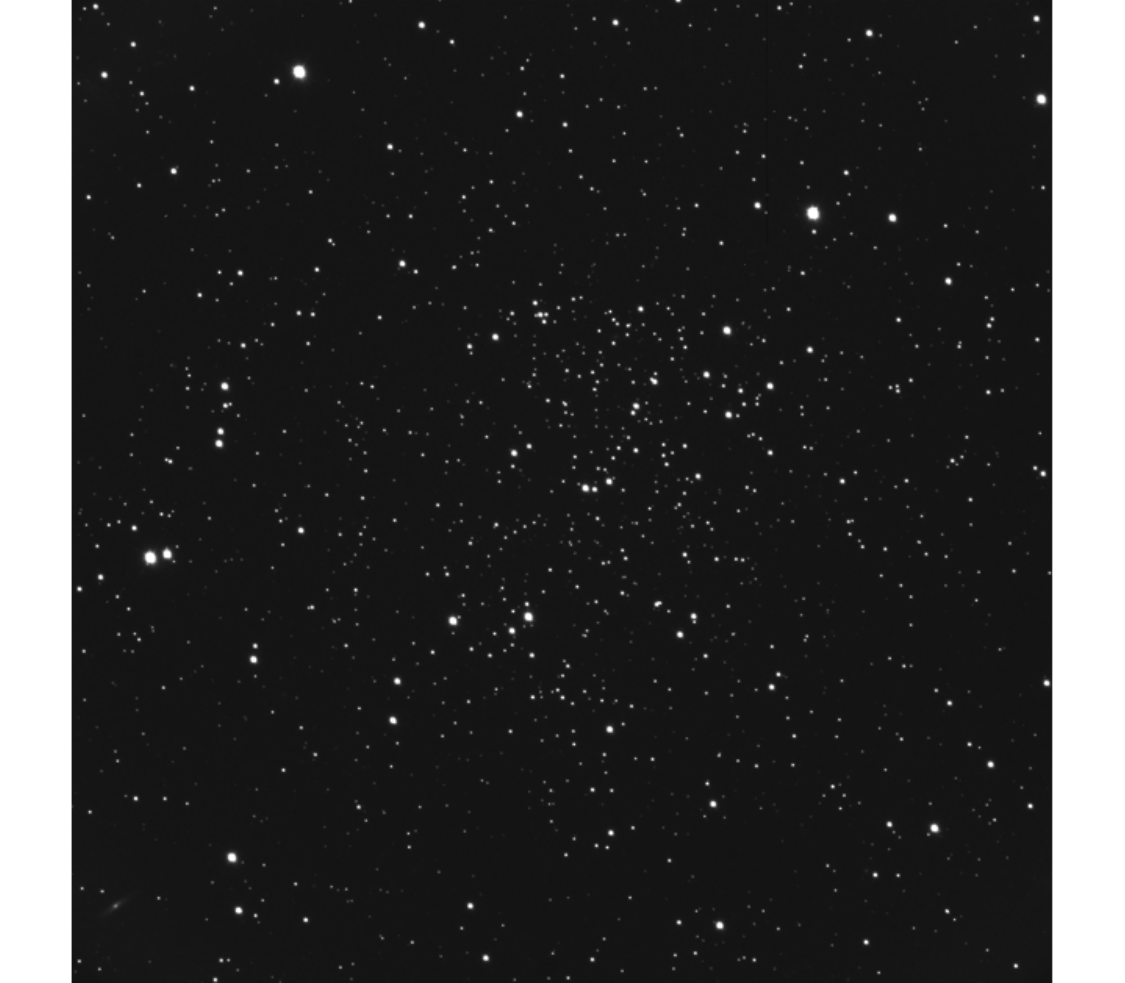}
\includegraphics[width=5cm,height=5cm]{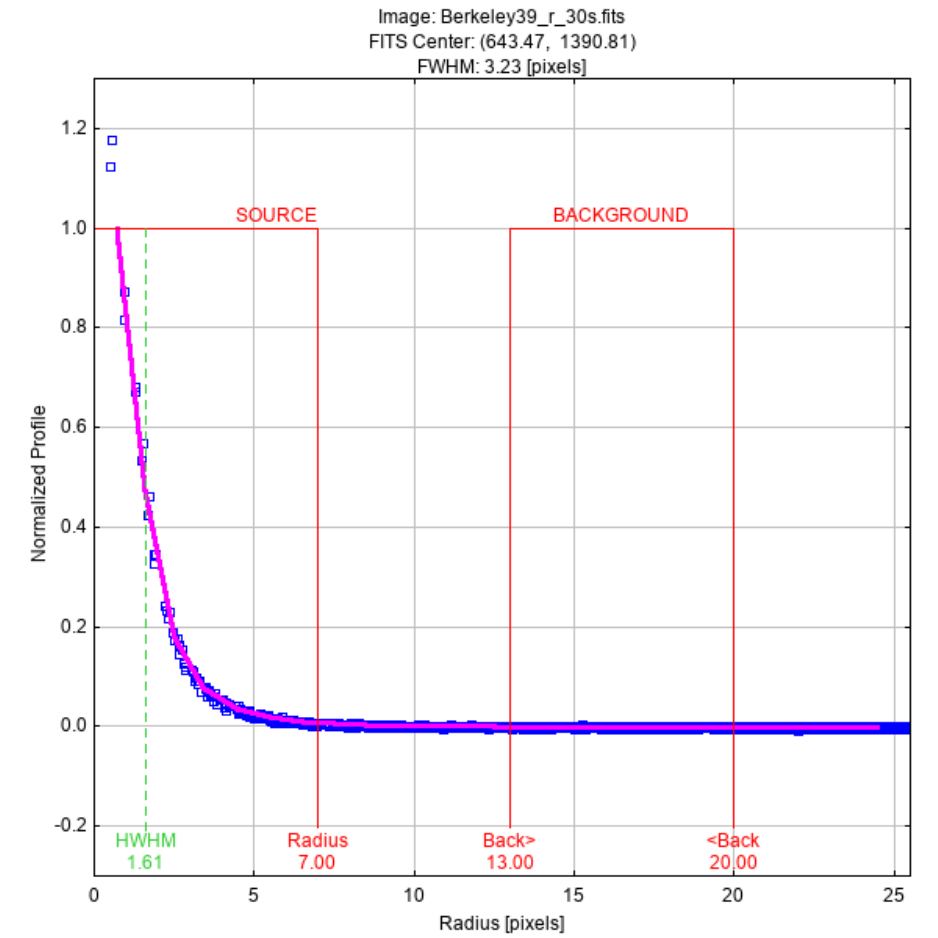}
\bigskip
\begin{minipage}{12cm}
\caption{An 30-sec image of the open cluster  {Berkeley 39} acquired using FOC after SAT, in the r'-band. The pixel scale of the camera is 0.15" pixel$^{-1}$. The image represents the best sky condition, in which PSF is measured around 3.2 pixels, which corresponds to $\sim$ 0.5". The radial profile of a star  {is} shown in the right panel.  }
\label{fig:4}
\end{minipage}
\end{figure}

\noindent The instrument utilizes a 4K x 4K deep depletion back-illuminated ANDOR CCD detector with a pixel size of 15~$\mu m$.  {This CCD has high Quantum Efficiency of $\sim$90\% in 400-800 nm, and 50\% at 1000 nm}. By cooling the CCD to -100$^{\circ}$C , the dark current is minimized to about 0.0008 electrons pix$^{-1}$ sec$^{-1}$. FOC's read noise is approximately 2.1 e$^{-}$ @ 100 kHz read rate, and 5 e$^{-}$ @ 500 KHz read rate with a quantum efficiency $\sim$ 90\%. The Field-of-View of FOC is 10' x 10'. In the r' filter, FOC can detect a 21$^{st}$ magnitude star in 10-minutes exposure time. It also includes a neutral density filter to image bright stars without detector saturation. FOC is a versatile instrument applicable to a wide range of astronomical studies, from the Solar system to extragalactic astronomy. The images shown in figure~\ref{fig:2} and figure~\ref{fig:4} are acquired with FOC.  We found that the typical seeing in the winter season (Oct-Feb) is < 1.0", {and} the best seeing was $\sim$ 0.5 arcsec in r'-band. (See figure~\ref{fig:4}). {Since regular observations began with the FOC in January 2023, we have observed sub-1" to 1" seeing in $\sim$60\% of the observation time under clear sky conditions. This perhaps is due to combination of multiple factors, including the exceptional quality of the telescope optics, the effectiveness of the precise Active Optics system, and efficient thermal control within the telescope dome area. To achieve dome thermalization, we employ air conditioning units to cool the dome area during the daytime, while at night, we activate the exhaust fans to maintain optimal thermal conditions relative to the external ambient temperature.}

\section{ {PARAS-2 Instrument}}
\subsection{Instrument Overview}{\label{overview}}
\noindent PRL Advanced Radial-velocity All-sky Search-2 (PARAS-2; \citep{Chakraborty2018}) is the advanced version of the existing spectrograph, PARAS \citep{Chakraborty2014} on the PRL 1.2m telescope. It is the first highly stabilized spectrograph of its kind in the country and sits among a few around the world. It is a fiber-fed high-resolution spectrograph that uses the white pupil design configuration where the star image at the telescope focal plane is re-imaged at the slit position of the spectrograph. It covers the wavelength range of 380 nm to 690 nm in a single shot. The spectrograph consists of optical elements like R4 echelle grating, M1 \& M2 off-axis parabolic mirrors, Grism as cross disperser, Camera lens barrel, and a back-thinned CCD of 6144 X 6190 pixels housed in a customized Dewar {designed by PRL}. A combination of octagonal and circular fibers transfers the light from the telescope focal plane to the spectrograph pupil plane. We use the simultaneous calibration (using Uranium Argon Hollow Cathode Lamps (UAr HCL) ) technique for getting down the Radial Velocity (RV) precision to 50 cm s$^{-1}$ or better. The optical and opto-mechanical design optimization of the PARAS-2 is carried out in-house in PRL. The spectrograph works at R$\sim$ 107,000 and is coupled with the PRL 2.5m telescope. The Cassegrain unit of the spectrograph is attached to one of the side ports of the telescope, that have tip-tilt to correct for first-order atmospheric seeing.

\subsection{Optical Design of the Spectrograph}{\label{ods}}
\noindent  {The optical design of PARAS-2 uses  {a} white pupil design configuration {\citep{Baranne1988}}.} The R4 echelle grating of useful size 200 $\times$ 800 mm is used for primary diffraction and blazed at an angle of 76.2$^{\circ}$ and has a groove frequency of 31.6 lines mm$^{-1}$. This grating, acquired from Richardson Gratings, Newport Corporation. The spectrograph is operating near the Littrow condition. The echelle is given an out-of-plane ($\gamma$) angle of 0.45$^{\circ}$ to direct the focused beam from the M1 mirror away from the slit position. We utilize a single large Grism as a cross-disperser with an apex angle of 13.24$^{\circ}$. It results in inter-order separations of approximately 42 pixels at 3800 {\AA}, 140 pixels at 6900{\AA}. The grism is specially designed to make it less sensitive to the temperature variations as the grating at its exit face compensate for the variations caused by the prism {\citep{Chakraborty2018}}. For a comprehensive understanding of the optical ray trace and discussions on optimization of the optical design and glass selections for the grism and camera optics, refer to \citet{Chakraborty2018}. The optical design of the spectrograph design was done in-house at PRL in collaboration with SAC-ISRO, Ahmedabad, India. The fabrication of the off-axis parabolic mirrors, camera lenses, and cylindrical glass window was carried out by SESO-THALES, France. The M1 and M2 are collimators which made from a single parabola. Our measurements of the surface roughness and the reflectivity of M1, M2 and the flat mirror are within the specified requirements. The camera lens barrel comprises three lenses: L1, an aspheric singlet; L2, a triplet; and L3, a doublet. After fabrication at the SESO facility, these lenses were meticulously aligned and bonded together before being carefully inserted into the camera lens barrel.
\begin{figure}[!ht]
\centering
\includegraphics[width=15cm]{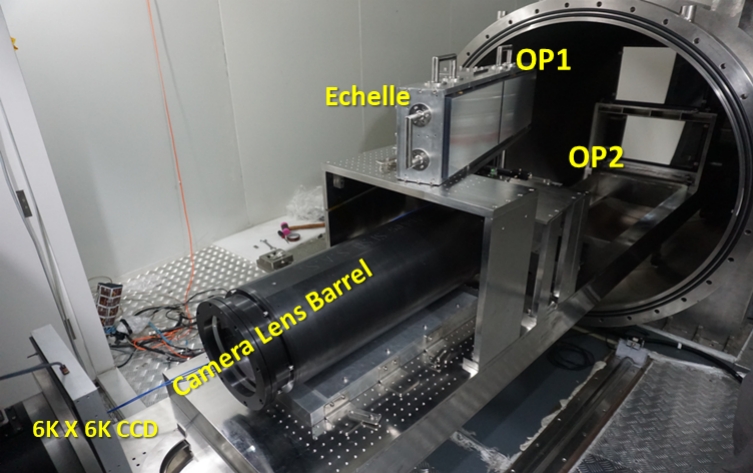}
\bigskip
\begin{minipage}{12cm}
\caption{The inside view of the vacuum chamber. The optical elements (echelle, Off-axis Parabolic mirrors (OP1 and OP2), the camera lens barrel and the detector) firmly kept over the optical bench.}
\label{fig:6}
\end{minipage}
\end{figure}

\subsection{Opto-mechanical components}
\noindent The opto-mechanical mounts for the echelle and grism, along with the vacuum chamber, optical bench, and CCD Dewar support, were all fabricated by Aditya High Vacuum Pvt. Ltd. in Ahmedabad, India, following their design at PRL. The other opto-mechanical holders for the mirrors and camera lenses were constructed and assembled at the SESO factory. These components were optimized for minimum deformations. The optical bench is made up in two stages where the top stage support the echelle and M1 mirror while the lower stage support the M2, grism, camera lens barrel and the CCD. The slit position with telecentric optics and fold mirror are kept just below the echelle. The opto-mechanical design ensures a consistent gravitational effect on all echelle grooves by aligning the gravity axis with them.

\subsection{CCD detector and Dewar}
\noindent The PARAS-2 detector is a back-thinned CCD detector with dimensions of 6144 $\times$ 6190 pixels, utilizing the E2V CCD231-C6 device. This CCD incorporates 15 $\mu m$ m square pixels and is equipped with an E2V "Astro-broadband" antireflection (AR) coating. This coating ensures high quantum efficiency (QE) across a broad range of wavelengths, specifically within the operating range of 380 nm to 690nm. The measured quantum efficiency of the CCD array is 95.4\% at 400 nm, 90.9\% at 500 nm, and 90.8\% at 650 nm. {This is an engineering grade CCD which has a 50-pixel wide bad column as shown in figure~\ref{fig:14}. In the near future, we will change this to scientific grade CCD (Grade Zero) with QE of greater than 90\% across the wavelength range of PARAS-2.}
\noindent The CCD Dewar mechanical design was customised at PRL and it was fabricated by InfraRed Laboratories, Arizona, US. It is designed to reduce the thermal gradient around the CCD chip. This dewar is kept outside of the vacuum chamber to avoid the local thermal concentration while filling up the LN2 (Liquid Nitrogen) tank of the Dewar. The design of the Dewar was customized for this requirement at PRL. Dewar is connected to the vacuum chamber through a metallic bellow which absorbs the excess vibrations, if any. This also helps in fine alignment of the spectrograph. 

\subsection{Integration and Alignment}
\noindent Initially, the spectrograph was assembled and aligned in a laboratory at PRL, Ahmedabad, India. The optical elements were positioned according to their respective theoretical locations, which were marked during the fabrication of the optical bench based on the Zemax design. This approach ensured that the optical components of the spectrograph were placed with a high degree of accuracy, typically within at a level of a few hundred $\mu m$  of their intended positions. Fine adjustments were made to the CCD Dewar position for precisely achieving the desired spot and resolution, ensuring optimal performance of the spectrograph.

\subsection{Fiber Optics of the spectrograph}
\noindent PARAS-2 is a fiber fed spectrograph where the light collected  from the telescope is transferred through a fiber and is fed into the spectrograph at its slit position. We are using two fibers from Polymicro Technologies, one as star fiber and other as calibration fiber. These fibers are a combination of octagonal and circular core of diameter 75 $\mu m$ . The total length of the fibers is $\sim$40m. The star fiber sees 1.5" on sky.  The spectrograph has no physical slit to avoid {light loss due to scattering and thus RV instabilities. The fiber tip image at the slit position itself acts as a slit. 

\subsubsection{ {Cassegrain unit and Fiber feed of PARAS-2}}\label{paras2_cass}
\noindent The Cassegrain unit of the spectrograph is attached with the side port 1 of the telescope {as shown in figure~\ref{fig:2}}. The M3 mirror at this port is equipped with tip-tilt support which mitigates the first order atmospheric seeing that helps in precise centering of the light falling onto the star fiber. It can operate up to 20 Hz speed. The tip-tilt corrections are sensitive up to 12$^{th}$ magnitude stars. The Cassegrain unit of the spectrograph is designed and developed within PRL. A schematic of PARAS-2 cassegrain unit is shown in figure~\ref{fig:8}.

\begin{figure}[!ht]
\centering
\includegraphics[width=16cm]{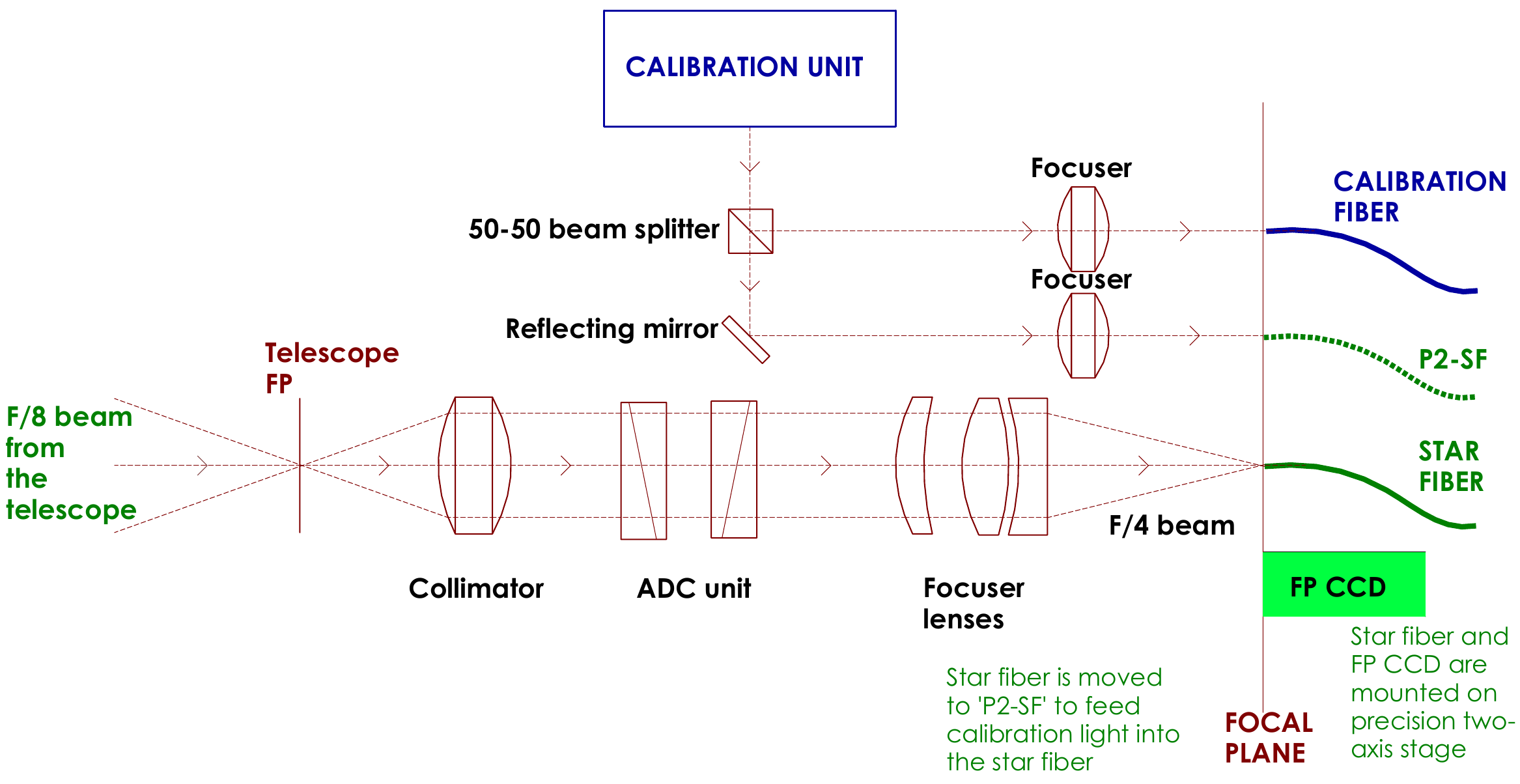}
\bigskip
\begin{minipage}{12cm}
\caption{{Schematic of the PARAS-2 Cassegrain unit (not to scale) showing how starlight
and calibration light are fed into the star and calibration fibers, respectively.}}
\label{fig:8}
\end{minipage}
\end{figure}

\noindent  {The F/8 beam from the telescope focal plane is converted to F/4 beam using a combination of collimator and focuser lenses. The star fiber tip is kept at the focal plane of this focal reducer. The fiber tip constantly centered at the F/4 beam focal plane for maximum signal during the actual scientific observations. Thus the structural design of the Cassegrain unit is aimed to be very rigid with minimum flexure during the telescope movement for scientific observations. Further the star fiber is supported over two axis support system which helps in maximizing the signal falling onto the fiber tip with fine adjustments. Two high-resolution linear actuators from Physik Instrumente (PI) have been used to develop this high-precision XY stage, which moves the star fiber tip and focal plane CCD camera (un-cooled Lodestar CCD from Starlight Xpress Ltd., FP CCD in figure~\ref{fig:8}) in the focal plane with a precision better than 3 $\mu m$ . We first focus the star on the focal plane CCD, and then the star fiber tip is moved by a fixed PI steps at the centroid of the star image. The calibration fiber tip is kept fixed. We have also employed an Atmospheric Dispersion Corrector (ADC) unit in the path of the starlight as shown in figure~\ref{fig:8} to mitigate the dispersive effects of the Earth's atmosphere. This addresses the differential atmospheric dispersion arising from the wavelength-dependent refractive index of the atmosphere.\\
\noindent The F/4 beam coming out of the fiber tip is converted back into F/8 using telecentric optics. The telecentric optics contain two achromatic doublets, which effectively gives a focal ratio of F/8. The F/8 beam focuses at the slit position of the spectrograph. The size of the fiber tip image is the effective slit width. The set of collimator and focuser lenses lenses as discussed earlier and the telecentric optics are designed in-house at PRL and manufactured by Luma Optics, India. }

\subsection{Pressure \& Temperature control}

\noindent The optical elements of PARAS-2 exhibit sensitivity to pressure and temperature variations. In RV measurements, a mere 1 mbar change in pressure or 1 $^{\circ}$C change in temperature can introduce RV variations of the order of 100 m s$^{-1}$ \citep{Pepe2002}. To attain the desired sub-m s$^{-1}$ ($\leq$50 cm s$^{-1}$) RV precision, the spectrograph's optics are housed within a vacuum chamber that in turns kept inside concentric cubicle chambers which are insulated and thermally stabilized \citep[figure~10b]{Chakraborty2018}. The vacuum chamber for PARAS-2 is designed in-house at PRL and manufactured by Aditya High Vacuum Pvt. Ltd. (AHV), India. Rigorous testing confirms its ability to reach a very low pressure of an order of 5e-06 mbar and stabilizing at 5e-03 mbar $\sim$ 18 hours, which is much longer than an observational night. The internal pressure of the vacuum chamber remains constant at 0.005 $\pm$ 0.0005 mbar during scientific observations. 
\begin{figure}[!ht]
\centering
\includegraphics[width=16cm]{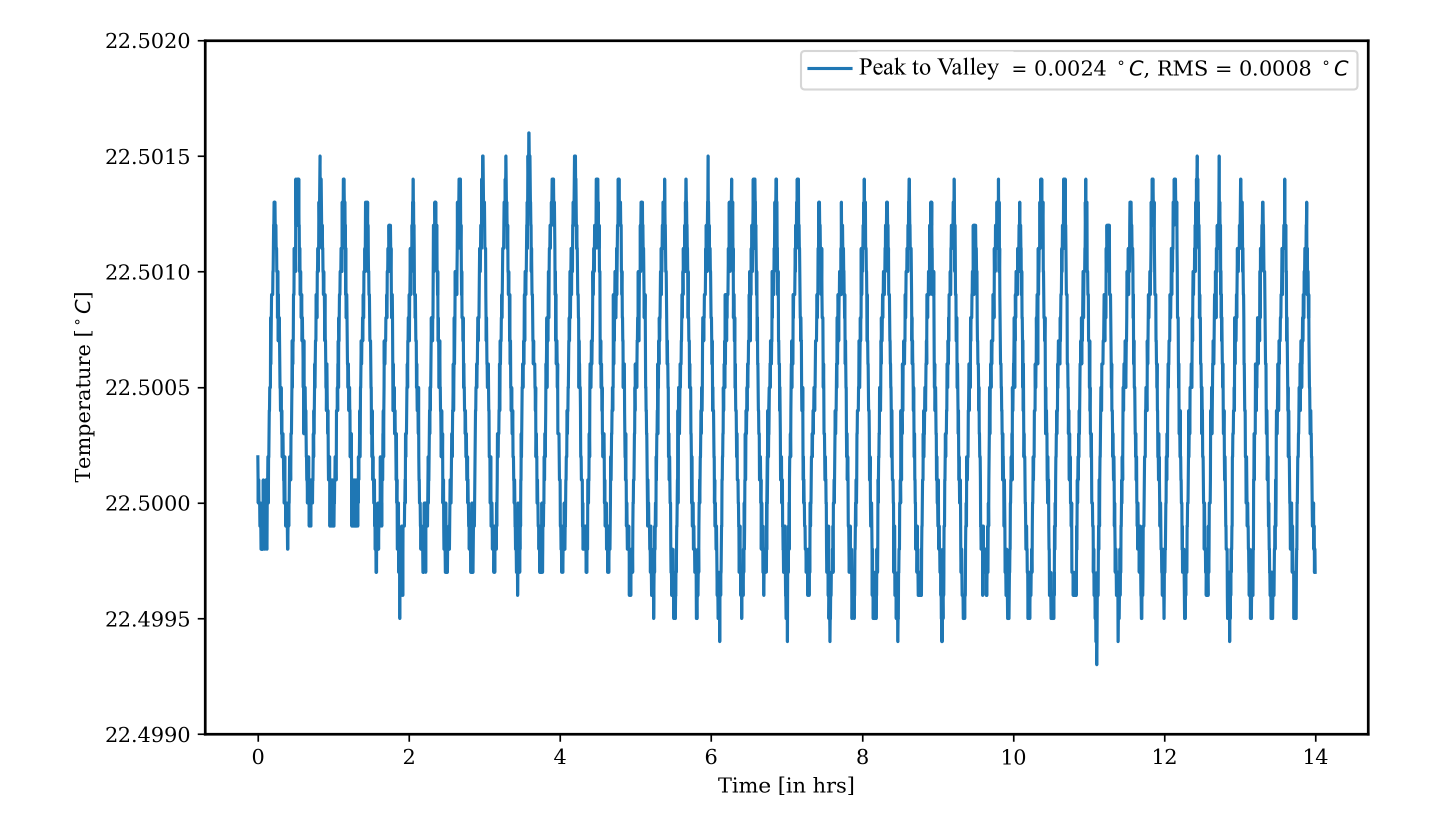}
\bigskip
\begin{minipage}{12cm}
\caption{Temperature variation over one typical observing night (Setpoint 22.5$^{\circ}$C), peak to valley variation is 0.0024$^{\circ}$C, and RMS variations is 0.0008$^{\circ}$C}
\label{fig:13}
\end{minipage}
\end{figure}

\noindent To ensure the temperature stability of the spectrograph, it is housed within two concentric cuboidal chambers that are thermally controlled. The setup includes an outer chamber and an inner chamber. The spectrograph is positioned on a reinforced concrete pier specifically designed for vibration isolation, ensuring isolation from all sides of the observatory building. Insulation is incorporated at the interface between the spectrograph and the pier. The walls of both chambers are constructed using highly insulated 60mm thick polyurethane foam (PUF) material.

\noindent Temperature control of the outer chamber is accomplished by utilizing a commercially available PID controller that employs heating and cooling mechanisms to maintain a temperature of 18$^{\circ}$C with a precision of 0.5$^{\circ}$C. For the spectrograph itself, an in-house developed temperature controller is employed to maintain a precise temperature of 22.5$^{\circ}$C with a precision of 0.001$^{\circ}$C. This comprehensive thermal regulation system ensures the stable temperature environment required for optimal performance of the PARAS-2 spectrograph as shown in figure~\ref{fig:13}.

\subsection{Spectrograph Format and Efficiency}

\noindent The combination of the R4 echelle and a large single 6144 $\times$ 6190 pixels CCD format (92 $\times$ 92 mm; pixel size of 15 $\mu m$ m) allows for coverage of a wide wavelength range from 3800 to 6900 {\AA } in a single shot, achieving a resolution of $\sim$ 107,000. The use of a Grism as a cross-disperser results in the spectral format having increasing order separation with increasing wavelength as discussed in section \ref{ods}. For an order, the separation between the two fibers is 22 pixels on the CCD detector. The two optical fibers which fed the light to the spectrograph are  called star-fiber or A-fiber and the other one is called Calibration-fiber or B-fiber. In figure~\ref{fig:14}, we are showing the spectrum of the Uranium-Argon Hollow cathode lamp with single fiber illumination (calibration-fiber). However, in general operation, both the fibers are illuminated for the simultaneous wavelength calibration. 
\begin{figure}[!ht]
\centering
\includegraphics[width=16cm]{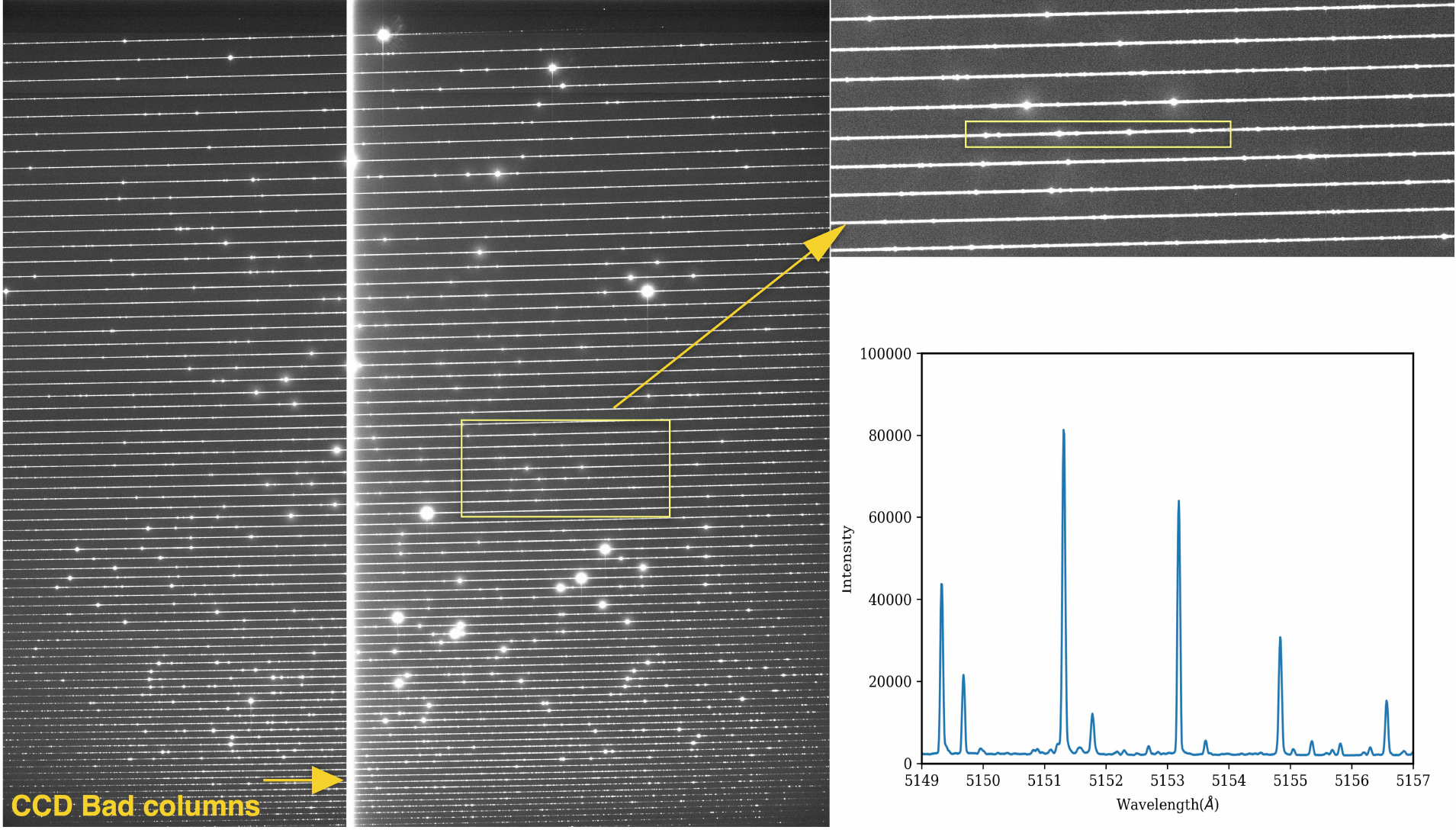}
\bigskip
\begin{minipage}{12cm}
\caption{The PARAS-2 echellogram (left) along with extracted UAr spectrum (right-bottom). There is a strip of 50 pixel wide bad column across the whole CCD, can be seen here. Currently, the CCD chip is an engineering grade chip, in near future we will change this
to scientific grade CCD chip.}
\label{fig:14}
\end{minipage}
\end{figure}

\subsection{Instrumental Stability}\label{wcis}

\noindent For testing purpose, we acquired a series of sequential UAr-UAr frames, in which both the fibers illuminated by the UAr HCL. The spectra from both the fibers were extracted using the existing PARAS pipeline \citep{Chakraborty2014} by manually adjusting few parameters. The wavelength calibration of these spectra is done using the Uranium linelist of \citep{Ulines2021}. These lines again used for calculating the instrumental drift. We created a binary mask of these Uranium lines, also known as template spectrum. The cross-correlation function (CCF) is calculated for a shifting uranium mask against each spectral order of acquired UAr spectrum, and the CCFs are summed in order to fit a Gaussian peak to the true drift value of the image. Here we are showing the absolute and relative instrumental fiber drifts in figure~\ref{fig:16}. We have consistently observed the dispersion in the inter-fiber drift to be 30-50 cm s$^{-1}$ which clearly demonstrates that PARAS-2 is capable of reaching precision in the order of sub-m s$^{-1}$.

\begin{figure}[!ht]
\centering
\includegraphics[width=14cm]{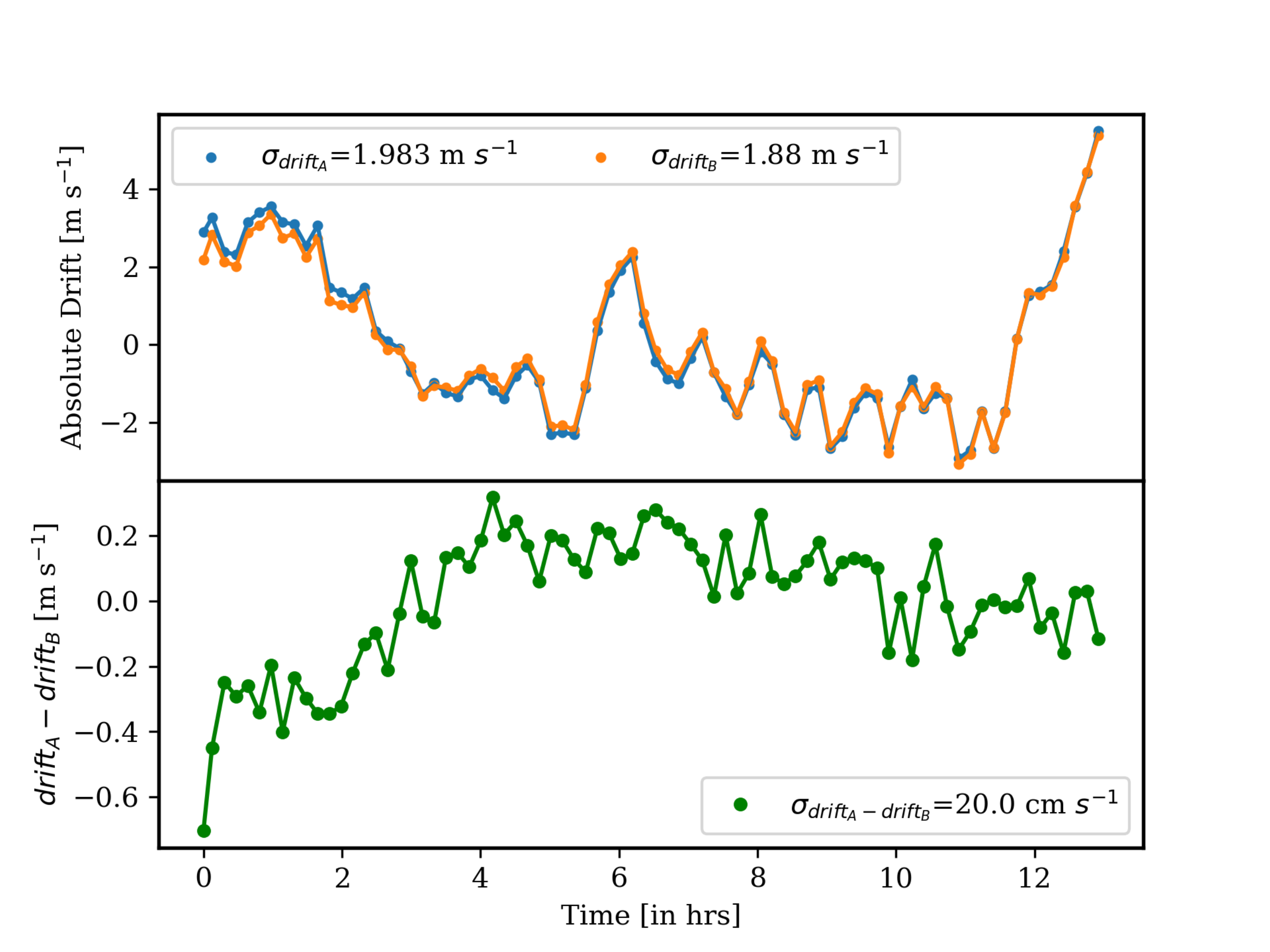}
\bigskip
\begin{minipage}{12cm}
\caption{Upper Panel:- The figure shows the absolute drift in A and B fiber with respect to time. The dispersion in absolute drift is $\sim$2 m s$^{-1}$ in both fibers. Lower Panel:-The figure shows the inter-fiber drift between the A and B fiber for the course of 12 hrs in m s$^{-1}$, measured using UAr HCL's spectra acquired in both the fibers, The dispersion of inter-fiber drift is 20 cm s$^{-1}$ for the mentioned duration.}
\label{fig:16}
\end{minipage}
\end{figure}

\section{Conclusion and Future works}
\noindent Regular on-sky observations with the 2.5m telescope have begun from January, 2023. This includes photometric observations of transient events like novae, supernovae, and star-forming regions using FOC.  On-sky observations with PARAS-2 have also begun. Although PARAS-2 is intrinsically stable down to 30 cm s$^{-1}$ (section \ref{wcis}), for on-sky measurements of stellar RVs, the non-uniform illumination of the of the star light at the spectrograph pupil plane remains in spite of using octagonal fibers, which necessitates installation of a double scrambler in the optical fiber path \citep{scrambler_main1992,Scrambler_hpf2015}.  At present, without the scrambler installed, on-sky RV precision on bright RV standard stars are typically around 1 to 3 m s-1. The double scrambler for PARAS-2 has been designed and currently under fabrication process. It is expected to be installed by the end of 2023. We expect on-sky RV precision on bright RV standard stars to be close to the intrinsic limit of the instrument (30-50 cm s$^{-1}$).   We are also in the process of installing the Fabry-Perot etalon for the precise wavelength calibration for the PARAS-2 \citep{FP_harps2010,FP_carmenes2018}. We will also replace the engineering grade 6K $\times$ 6K CCD with the science grade CCD for PARAS-2 by the end of 2023.\\
PRL is also developing various other backend instruments for the 2.5m telescope, such as near infrared spectrograph and polarimeter (NISP, \citet{NISP2020, NISP_1_2020}), low-resolution spectrograph (LRS) and an echelle spectro-polarimeter (MFOSC-EP \& its prototype, \citep{MFOSCEP2022}) which will see the first light in the upcoming 1-2 years for various astrophysical studies.  {A detailed paper on the performance of the spectrograph and its various sub-systems is under preparation.}

\begin{acknowledgments}
Authors would like to express their heartfelt gratitude to all those who have contributed to the development of the telescope and the PARAS-2 spectrograph. Firstly, we extend our sincere appreciation to the Director of Physical Research Laboratory (PRL) for his invaluable support and resources provided throughout the project. We are very thankful to the Department of Space, India, for providing the funding for this project. We would also like to extend our thanks to the team at AMOS Belgium, especially Mr. Olivier Pirnay, for their expertise in the development of the telescope. Their technical knowledge has played a crucial role {in the success of the 2.5m telescope project.} Furthermore, we would like to acknowledge the Mount Abu Observatory staff for their unwavering support and assistance during the telescope\textquotesingle s and PARAS-2 spectrograph\textquotesingle s integration and testing phase. Their commitment and hard work have been instrumental in ensuring its smooth functioning. We express our gratitude Prof. Andrew Szentgyorgyi  (senior professor, SAO, Harvard) and {Prof.} Francesco Pepe (Director, Geneva Observatory) for giving us valuable inputs during the design and testing phase of PARAS-2. We also extend our special thanks to {Mr.} Jaimin Desai (Engineer G, SAC-ISRO), {Mr.} D. Subrahmanyam (Retired Deputy Director, SAC-ISRO, Ahmedabad) and, Dr. DRMS Samudraiah (Retired, Deputy Director, SAC-ISRO, Ahmedabad), for their inputs during the telescope design phase of the project. We would also like to thank Mr. Julien Fourez and his team at SESO THALES, who have helped in further optimization of the camera lens barrel with the melt data. RS would like to thank Mr. Sanjay Baliwal for his help in acquisition and analysis of the PARAS-2 in the testing phase. KB likes to extend his gratitude to Dr. Vankat Ramani and his team from AHV for his frequent visits to Mount Abu Observatory for their help in PARAS-2 vacuum chamber. {AC would like to thank Dr. Mudit Srivastava for his help in the procurement of telescope accesories like chiller, compressor and various other items. Last but not least, we extend our gratitude to all other individuals and teams specially PRL workshop members involved in this project. Your contributions, whether big or small, have been vital in the realization of this project. We are sincerely grateful for everyone's efforts and commitment to this endeavour, which will undoubtedly contribute to advancements in astronomical research.} 
\end{acknowledgments}

\bibliographystyle{apalike}
\bibliography{references}

\begin{thebibliography}{}

\bibitem[{Anandarao} and {Chakraborty}, 2010]{Anandarao2010}
{Anandarao}, B.~G. and {Chakraborty}, A. (2010).
\newblock {PRL Mt. Abu Observatory: New initiatives}.
\newblock In {\em Astronomical Society of India Conference Series}, volume~1 of
  {\em Astronomical Society of India Conference Series}, page 211.

\bibitem[{Ashok} and {Banerjee}, 2002]{site_1}
{Ashok}, N.~M. and {Banerjee}, D.~P.~K. (2002).
\newblock {Limited Seeing Measurements at Mount Abu Infrared Observatory}.
\newblock {\em BASI}, 30:851--858.

\bibitem[{Ashok} and {Banerjee}, 2003]{AshokN2003}
{Ashok}, N.~M. and {Banerjee}, D.~P.~K. (2003).
\newblock {JHK Spectroscopy of the Enigmatic Variable V445 Puppis}.
\newblock {\em BASI}, 31:343--344.

\bibitem[{Banerjee} and {Ashok}, 2002]{Banerjee2002}
{Banerjee}, D.~P.~K. and {Ashok}, N.~M. (2002).
\newblock {Near infra-red spectroscopy of V838 Monocerotis}.
\newblock {\em A\&A}, 395:161--167.

\bibitem[{Baranne}, 1988]{Baranne1988}
{Baranne}, A. (1988).
\newblock {White Pupil Story or Evolution of a Spectrographic Mounting}.
\newblock In {\em Very Large Telescopes and their Instrumentation, Vol. 2},
  volume~30 of {\em European Southern Observatory Conference and Workshop
  Proceedings}, page 1195.

\bibitem[{Bastin} et~al., 2022]{Bastin2022}
{Bastin}, C., {Lousberg}, G.~P., {Pirnay}, O., {Adam}, C., {Albart}, P., {De
  Ville}, J., {Feutry}, A., {Flebus}, C., {Fontana}, N., {Gabriel}, E.,
  {Lanotte}, A.~A., {Lemagne}, F., {Leseur}, T., {M{\'e}ant}, L., {Ninane}, N.,
  {Orban}, S., {Tortolani}, J.-M., and {Verspecht}, J. (2022).
\newblock {Mount Abu 2.5m Telescope: first light and performance assessment}.
\newblock In {Angeli}, G.~Z. and {Dierickx}, P., editors, {\em Modeling,
  Systems Engineering, and Project Management for Astronomy X}, volume 12187 of
  {\em SPIE}, page 121870H.

\bibitem[{Benz} et~al., 2021]{Benz2020}
{Benz}, W., {Broeg}, C., {Fortier}, A., {Rando}, N., {Beck}, T., {Beck}, M.,
  {Queloz}, D., {Ehrenreich}, D., {Maxted}, P.~F.~L., {Isaak}, K.~G., {Billot},
  N., {Alibert}, Y., {Alonso}, R., {Ant{\'o}nio}, C., and {Asquier}, J.~e.
  (2021).
\newblock {The CHEOPS mission}.
\newblock {\em ExA}, 51(1):109--151.

\bibitem[{Chakraborty} et~al., 2014]{Chakraborty2014}
{Chakraborty}, A., {Mahadevan}, S., {Roy}, A., {Dixit}, V., {Richardson},
  E.~H., {Dongre}, V., {Pathan}, F.~M., {Chaturvedi}, P., {Shah}, V., {Ubale},
  G.~P., and {Anandarao}, B.~G. (2014).
\newblock {The PRL Stabilized High-Resolution Echelle Fiber-fed Spectrograph:
  Instrument Description and First Radial Velocity Results}.
\newblock {\em PASP}, 126(936):133.

\bibitem[{Chakraborty} et~al., 2008]{Chakraborty2008}
{Chakraborty}, A., {Richardson}, E.~H., and {Mahadevan}, S. (2008).
\newblock {PRL advanced radial-velocity all-sky search (PARAS): an efficient
  fiber-fed spectrograph for planet searches}.
\newblock In {McLean}, I.~S. and {Casali}, M.~M., editors, {\em Ground-based
  and Airborne Instrumentation for Astronomy II}, volume 7014 of {\em SPIE},
  page 70144G.

\bibitem[{Chakraborty} et~al., 2018]{Chakraborty2018}
{Chakraborty}, A., {Thapa}, N., {Kumar}, K., {Neelam}, P. J.~S.~S.~V.,
  {Sharma}, R., and {Roy}, A. (2018).
\newblock {PARAS-2 precision radial velocimeter: optical and mechanical design
  of a fiber-fed high resolution spectrograph under vacuum and temperature
  control}.
\newblock In {Evans}, C.~J., {Simard}, L., and {Takami}, H., editors, {\em
  Ground-based and Airborne Instrumentation for Astronomy VII}, volume 10702 of
  {\em SPIE}, page 107026G.

\bibitem[{Cort{\'e}s-Contreras} et~al., 2020]{GTC_filters}
{Cort{\'e}s-Contreras}, M., {Bouy}, H., {Solano}, E., {Mahlke}, M.,
  {Jim{\'e}nez-Esteban}, F., {Alacid}, J.~M., and {Rodrigo}, C. (2020).
\newblock {The Gran Telescopio Canarias OSIRIS broad-band first data release}.
\newblock {\em MNRAS}, 491(1):129--152.

\bibitem[{Deshpande}, 1995]{telescope}
{Deshpande}, M.~R. (1995).
\newblock {A brief report on the Infrared Telescope at Gurushikhar, MT Abu}.
\newblock {\em BASI}, 23:13.

\bibitem[{Halverson} et~al., 2015]{Scrambler_hpf2015}
{Halverson}, S., {Roy}, A., {Mahadevan}, S., {Ramsey}, L., {Levi}, E.,
  {Schwab}, C., {Hearty}, F., and {MacDonald}, N. (2015).
\newblock {An Efficient, Compact, and Versatile Fiber Double Scrambler for High
  Precision Radial Velocity Instruments}.
\newblock {\em ApJ}, 806(1):61.

\bibitem[Howell et~al., 2014]{Howell_2014}
Howell, S.~B., Sobeck, C., Haas, M., Still, M., Barclay, T., Mullally, F.,
  Troeltzsch, J., Aigrain, S., Bryson, S.~T., Caldwell, D., Chaplin, W.~J.,
  Cochran, W.~D., Huber, D., Marcy, G.~W., Miglio, A., Najita, J.~R., Smith,
  M., Twicken, J.~D., and Fortney, J.~J. (2014).
\newblock The k2 mission: Characterization and early results.
\newblock {\em PASP}, 126(938):398.

\bibitem[{Hunter} and {Ramsey}, 1992]{scrambler_main1992}
{Hunter}, T.~R. and {Ramsey}, L.~W. (1992).
\newblock {Scrambling Properties of Optical Fibers and the Performance of a
  Double Scrambler}.
\newblock {\em PASP}, 104:1244.

\bibitem[{Kasarla} et~al., 2020]{NISP_1_2020}
{Kasarla}, P.~K., {Patwal}, P.~S., {Adalja}, H. K.~L., {Mathur}, S.~N.,
  {Sarkar}, D.~R., {Singh}, A., {Rai}, A., {Prajapati}, P.~V., {Naik}, S.,
  {Shah}, A.~B., {Ganesh}, S., and {Baliyan}, K.~S. (2020).
\newblock {Mechanical aspects of near infrared imager spectrometer and
  polarimeter}.
\newblock In {Evans}, C.~J., {Bryant}, J.~J., and {Motohara}, K., editors, {\em
  Ground-based and Airborne Instrumentation for Astronomy VIII}, volume 11447
  of {\em SPIE}, page 114476U.

\bibitem[{Kumar} et~al., 2022]{MFOSCEP2022}
{Kumar}, V., {Srivastava}, M.~K., {Dixit}, V., {Mistry}, B., {Lad}, K.,
  {Patel}, A., and {Rajpurohit}, A.~S. (2022).
\newblock {Designs of Mt. Abu faint object spectrograph and camera - echelle
  polarimeter (M-FOSC-EP) and its prototype: spectro-polarimeters for PRL 1.2m
  and 2.5m Mt. Abu Telescopes, India}.
\newblock In {Evans}, C.~J., {Bryant}, J.~J., and {Motohara}, K., editors, {\em
  Ground-based and Airborne Instrumentation for Astronomy IX}, volume 12184 of
  {\em SPIE}, page 121845B.

\bibitem[Mahadevan et~al., 2012]{Mahadevan2012}
Mahadevan, S., Ramsey, L., Bender, C., Terrien, R., Wright, J., Halverson, S.,
  Hearty, F., Nelson, M., Burton, A., Redman, S., Osterman, S., Diddams, S.,
  Kasting, J., Endl, M., and Deshpande, R. (2012).
\newblock The habitable-zone planet finder: A stabilized fiber-fed nir
  spectrograph for the hobby-eberly telescope.
\newblock {\em SPIE}, 8446.

\bibitem[{Mahadevan} et~al., 2014]{Mahadevan2014}
{Mahadevan}, S., {Ramsey}, L.~W., {Terrien}, R., {Halverson}, S., {Roy}, A.,
  {Hearty}, F., {Levi}, E., {Stefansson}, G.~K., {Robertson}, P., {Bender}, C.,
  {Schwab}, C., and {Nelson}, M. (2014).
\newblock {The Habitable-zone Planet Finder: A status update on the development
  of a stabilized fiber-fed near-infrared spectrograph for the for the
  Hobby-Eberly telescope}.
\newblock In {Ramsay}, S.~K., {McLean}, I.~S., and {Takami}, H., editors, {\em
  Ground-based and Airborne Instrumentation for Astronomy V}, volume 9147 of
  {\em SPIE}, page 91471G.

\bibitem[{Mayor} et~al., 2003]{Mayor2003}
{Mayor}, M., {Pepe}, F., {Queloz}, D., {Bouchy}, F., {Rupprecht}, G., {Lo
  Curto}, G., {Avila}, G., {Benz}, W., {Bertaux}, J.~L., {Bonfils}, X., {Dall},
  T., {Dekker}, H., {Delabre}, B., {Eckert}, W., {Fleury}, M., {Gilliotte}, A.,
  {Gojak}, D., {Guzman}, J.~C., {Kohler}, D., {Lizon}, J.~L., {Longinotti}, A.,
  {Lovis}, C., {Megevand}, D., {Pasquini}, L., {Reyes}, J., {Sivan}, J.~P.,
  {Sosnowska}, D., {Soto}, R., {Udry}, S., {van Kesteren}, A., {Weber}, L., and
  {Weilenmann}, U. (2003).
\newblock {Setting New Standards with HARPS}.
\newblock {\em Msngr}, 114:20--24.

\bibitem[{Pepe} et~al., 2002]{Pepe2002}
{Pepe}, F., {Mayor}, M., {Rupprecht}, G., {Avila}, G., {Ballester}, P.,
  {Beckers}, J.~L., {Benz}, W., {Bertaux}, J.~L., {Bouchy}, F., {Buzzoni}, B.,
  {Cavadore}, C., {Deiries}, S., {Dekker}, H., {Delabre}, B., {D'Odorico}, S.,
  {Eckert}, W., {Fischer}, J., {Fleury}, M., {George}, M., {Gilliotte}, A.,
  {Gojak}, D., {Guzman}, J.~C., {Koch}, F., {Kohler}, D., {Kotzlowski}, H.,
  {Lacroix}, D., {Le Merrer}, J., {Lizon}, J.~L., {Lo Curto}, G., {Longinotti},
  A., {Megevand}, D., {Pasquini}, L., {Petitpas}, P., {Pichard}, M., {Queloz},
  D., {Reyes}, J., {Richaud}, P., {Sivan}, J.~P., {Sosnowska}, D., {Soto}, R.,
  {Udry}, S., {Ureta}, E., {van Kesteren}, A., {Weber}, L., {Weilenmann}, U.,
  {Wicenec}, A., {Wieland}, G., {Christensen-Dalsgaard}, J., {Dravins}, D.,
  {Hatzes}, A., {K{\"u}rster}, M., {Paresce}, F., and {Penny}, A. (2002).
\newblock {HARPS: ESO's coming planet searcher. Chasing exoplanets with the La
  Silla 3.6-m telescope}.
\newblock {\em Msngr}, 110:9--14.

\bibitem[{Pepe} et~al., 2010]{Pepe2010}
{Pepe}, F.~A., {Cristiani}, S., {Rebolo Lopez}, R., {Santos}, N.~C., {Amorim},
  A., {Avila}, G., {Benz}, W., {Bonifacio}, P., {Cabral}, A., {Carvas}, P.,
  {Cirami}, R., {Coelho}, J., {Comari}, M., and {et al.} (2010).
\newblock {ESPRESSO: the Echelle spectrograph for rocky exoplanets and stable
  spectroscopic observations}.
\newblock In {McLean}, I.~S., {Ramsay}, S.~K., and {Takami}, H., editors, {\em
  Ground-based and Airborne Instrumentation for Astronomy III}, volume 7735 of
  {\em SPIE}, page 77350F.

\bibitem[{Pirnay} et~al., 2018]{PirnayO2018}
{Pirnay}, O., {Lousberg}, G.~P., {Lanotte}, A., {Gabriel}, E., {Tortolani},
  J.-M., {Fontana}, N., and {Orban}, S. (2018).
\newblock {Mt ABU 2.5m Telescope: design and fabrication}.
\newblock In {Marshall}, H.~K. and {Spyromilio}, J., editors, {\em Ground-based
  and Airborne Telescopes VII}, volume 10700 of {\em SPIE}, page 107005S.

\bibitem[{Quirrenbach} et~al., 2016]{Quirrenbach2016}
{Quirrenbach}, A., {Amado}, P.~J., {Caballero}, J.~A., {Mundt}, R., {Reiners},
  A., {Ribas}, I., {Seifert}, W., {Abril}, M., {Aceituno}, J.,
  {Alonso-Floriano}, F.~J., {Anwand-Heerwart}, H., {Azzaro}, M., {Bauer}, F.,
  {Barrado}, D., {Becerril}, S., {Bejar}, V.~J.~S., {Benitez}, D., and {et al.}
  (2016).
\newblock {CARMENES: an overview six months after first light}.
\newblock In {Evans}, C.~J., {Simard}, L., and {Takami}, H., editors, {\em
  Ground-based and Airborne Instrumentation for Astronomy VI}, volume 9908 of
  {\em SPIE}, page 990812.

\bibitem[{Rai} et~al., 2020]{NISP2020}
{Rai}, A., {Ganesh}, S., {Paul}, S.~K., {Kasarla}, P.~K., {Prajapati}, P.~V.,
  {Sarkar}, D.~R., {Singh}, A., {Patwal}, P.~S., {Adalja}, H. K.~L., {Mathur},
  S.~N., {Naik}, S., {Shah}, A.~B., and {Baliyan}, K.~S. (2020).
\newblock {Optical aspects of Near-Infrared Imager Spectrometer and Polarimeter
  instrument (NISP)}.
\newblock In {Evans}, C.~J., {Bryant}, J.~J., and {Motohara}, K., editors, {\em
  Ground-based and Airborne Instrumentation for Astronomy VIII}, volume 11447
  of {\em SPIE}, page 1144765.

\bibitem[{Rauer} et~al., 2014]{Rauer2014}
{Rauer}, H., {Catala}, C., {Aerts}, C., {Appourchaux}, T., {Benz}, W.,
  {Brandeker}, A., {Christensen-Dalsgaard}, J., {Deleuil}, M., {Gizon}, L.,
  {Goupil}, M.~J., {G{\"u}del}, M., {Janot-Pacheco}, E., {Mas-Hesse}, M.,
  {Pagano}, I., and {Piotto}, G.~e. (2014).
\newblock {The PLATO 2.0 mission}.
\newblock {\em ExA}, 38(1-2):249--330.

\bibitem[{Ricker} et~al., 2015]{Ricker2015}
{Ricker}, G.~R., {Winn}, J.~N., {Vanderspek}, R., {Latham}, D.~W., {Bakos},
  G.~{\'A}., {Bean}, J.~L., {Berta-Thompson}, Z.~K., {Brown}, T.~M.,
  {Buchhave}, L., {Butler}, N.~R., {Butler}, R.~P., {Chaplin}, W.~J.,
  {Charbonneau}, D., and {et al.} (2015).
\newblock {Transiting Exoplanet Survey Satellite (TESS)}.
\newblock {\em JATIS}, 1:014003.

\bibitem[{Sapru} et~al., 1993]{site_2}
{Sapru}, M.~L., {Bhat}, C.~L., {Kaul}, M.~K., {Kaul}, S.~K., {Kaul}, S.~R.,
  {Dhar}, V.~K., {Kaul}, R.~K., {Rannot}, R.~C., and {Tickoo}, A.~K. (1993).
\newblock {Site-selection studies at Gurushikar, Mt. Abu for projet ``GRACE''.}
\newblock {\em BASI}, 21(3-4):515--518.

\bibitem[{Sch{\"a}fer} et~al., 2018]{FP_carmenes2018}
{Sch{\"a}fer}, S., {Guenther}, E.~W., {Reiners}, A., {Winkler}, J., {Pluto},
  M., and {Schiller}, J. (2018).
\newblock {Two Fabry-P{\'e}rots and two calibration units for CARMENES}.
\newblock In {Evans}, C.~J., {Simard}, L., and {Takami}, H., editors, {\em
  Ground-based and Airborne Instrumentation for Astronomy VII}, volume 10702 of
  {\em SPIE}, page 1070276.

\bibitem[{Schwab} et~al., 2016]{Schwab2016}
{Schwab}, C., {Rakich}, A., {Gong}, Q., {Mahadevan}, S., {Halverson}, S.~P.,
  {Roy}, A., {Terrien}, R.~C., {Robertson}, P.~M., {Hearty}, F.~R., {Levi},
  E.~I., {Monson}, A.~J., {Wright}, J.~T., {McElwain}, M.~W., {Bender}, C.~F.,
  {Blake}, C.~H., {St{\"u}rmer}, J., {Gurevich}, Y.~V., {Chakraborty}, A., and
  {Ramsey}, L.~W. (2016).
\newblock {Design of NEID, an extreme precision Doppler spectrograph for WIYN}.
\newblock In {Evans}, C.~J., {Simard}, L., and {Takami}, H., editors, {\em
  Ground-based and Airborne Instrumentation for Astronomy VI}, volume 9908 of
  {\em SPIE}, page 99087H.

\bibitem[{Sharma} and {Chakraborty}, 2021]{Ulines2021}
{Sharma}, R. and {Chakraborty}, A. (2021).
\newblock {Precision wavelength calibration for radial velocity measurements
  using uranium lines between 3800 and 6900 {\r{A}}}.
\newblock {\em JATIS}, 7:038005.

\bibitem[{Wildi} et~al., 2010]{FP_harps2010}
{Wildi}, F., {Pepe}, F., {Chazelas}, B., {Lo Curto}, G., and {Lovis}, C.
  (2010).
\newblock {A Fabry-Perot calibrator of the HARPS radial velocity spectrograph:
  performance report}.
\newblock In {McLean}, I.~S., {Ramsay}, S.~K., and {Takami}, H., editors, {\em
  Ground-based and Airborne Instrumentation for Astronomy III}, volume 7735 of
  {\em SPIE}, page 77354X.

\end{thebibliography}

\end{document}